\documentclass[trackchanges,twocolumn]{aastex7}
\usepackage{amsmath}

\begin{document}

\title{Simulating Disky Broad Line Region Reverberation}

\author[0000-0001-9741-2703]{Mary Ogborn}
\email[show]{mko5309@psu.edu}
\affiliation{Department of Astronomy and Astrophysics and Institute for Gravitation and the Cosmos, The Pennsylvania State University, State College, PA 16802, USA}

\author[0000-0002-3719-940X]{Michael Eracleous}
\email[]{mxe17@psu.edu}
\affiliation{Department of Astronomy and Astrophysics and Institute for Gravitation and the Cosmos, The Pennsylvania State University, State College, PA 16802, USA}

\author[0000-0001-8557-2822]{Jessie C. Runnoe}
\email[]{jessie.c.runnoe@vanderbilt.edu}
\affil{Department of Physics \& Astronomy, Vanderbilt University, 2301 Vanderbilt Place, Nashville, TN 37235, USA}
\affil{Department of Life and Physical Sciences, Fisk University, 1000 17th Avenue N, Nashville, TN 37208, USA}

\author[0000-0002-1683-5198]{Massimo Dotti}
\email[]{massimo.dotti@unimib.it}
\affil{Università degli Studi di Milano-Bicocca, Piazza della Scienza 3, 20126 Milano, Italy}
\affil{INAF - Osservatorio Astronomico di Brera, via Brera 20, I-20121 Milano, Italy}
\affil{INFN, Sezione di Milano-Bicocca, Piazza della Scienza 3, I-20126 Milano, Italy}

\author[0000-0001-9806-4034]{Niana N. Mohammed}
\email[]{nnm5189@psu.edu}
\affiliation{Department of Astronomy and Astrophysics and Institute for Gravitation and the Cosmos, The Pennsylvania State University, State College, PA 16802, USA}

\author[0000-0001-7306-1830]{Collin M. Dabbieri}
\email[]{collin.m.dabbieri@vanderbilt.edu}
\affil{Department of Physics \& Astronomy, Vanderbilt University, 2301 Vanderbilt Place, Nashville, TN 37235, USA}

\begin{abstract}

Variability studies of the broad emission lines of Active Galactic Nuclei (AGNs) and quasars show stochastic radial velocity variations (i.e., fluctuations in the centroid of the line), ‘jitter’, on timescales of weeks to months. This jitter may be intrinsic as the broad-line emitting region (BLR) reverberates from the AGN continuum. There are also coordinated variations in the width of the broad emission lines and the luminosity of the central source (‘breathing’ or ‘anti-breathing’) which remain unexplained. These can be used as a tool for testing models of the BLR. We have constructed a pipeline to simulate a disk-like BLR geometry that reverberates in response to various chosen continuum light curves and produce synthetic emission line profiles. These profiles can then be characterized by measured shape parameters (centroid velocity shift, velocity dispersion, and Pearson skewness coefficient) and compared to observed time series of those same parameters. We have found that through our pipeline, we can recreate the velocity jitter at similar variations found in observations. The computational tools presented in this paper will also be applicable to case studies of quasars observed under the Sloan Digital Sky Survey V (SDSS-V) Black Hole Mapper reverberation mapping program. This paper is the first in a series of papers---in this paper, we present the model and pipeline, and in future papers, we will present applications.

\end{abstract}



\section{Introduction} 
\label{sec:intro}
The optical spectra of active galactic nuclei (AGN) and quasars often show broad emission lines \cite[]{khachikian74} that have widths on the order of $10^{3-4}$ km s$^{-1}$ \citep[e.g.,][]{peterson93}. The origin of these broad emission lines is the gaseous region correspondingly called the Broad-Line Region (BLR), which is photoionized by the high-energy continuum coming from the central engine of the AGN. Observations of the emission lines show the same variability pattern as the observed optical and UV continuum with a smaller amplitude and a time delay. This behavior is known as reverberation of the BLR. Reverberation mapping \cite[]{blandford82} can be used to probe the structure and kinematics of the BLR, and has obtained constraints on the extent of the BLR and provides most of the information currently known on BLR sizes. Only a few BLR geometries have been successfully spatially resolved using interferometric methods \citep[e.g.,][]{gravity20}, which further confirm the results of reverberation mapping. 

Many intensive reverberation mapping campaigns have been carried out, starting with efforts in the 1990s \citep[see][]{clavel91, korista95} and progressing to more modern efforts, such as the LAMP campaign \citep[see][]{bentz09}, AGN STORM and AGN STORM 2 \citep[see][]{derosa15, kara21}, the Sloan Digital Sky Surveys \citep[see][]{eisenstein11, blanton17, shen15}, the Supermassive Black Holes with High Accretion Rates in Active Galactic Nuclei program \citep[see][]{du14}, and the Seoul National University AGN Monitoring Project \citep[see][]{woo19}. As a result of such campaigns, a number of behaviors have been identified observationally in the reverberation signatures of the BLR. A particular behavior that is vital for obtaining black hole masses in these campaigns in that the BLR is ``virialized" \citep{peterson99, grier13, u22}, or that the gas in the BLR obeys the virial theorem \citep{peterson99, peterson00}. Another behavior observed is the ``radius-luminosity" relation -- a correlation between time lag and the continuum luminosity \citep[e.g.,][]{kaspi00, bentz13}. 

Other studies have shown that there is an anti-correlation between the luminosity of the AGN and the broad emission line widths, notably when examining the H$\beta$ line \citep[e.g.,][]{dexter19}. This effect, known as `breathing', can produce a narrower line profile by boosting the core's low velocity flux compared to the higher velocity wings of the profile. \cite{korista04} and \cite{goad14} attribute this behavior to the higher responsivity of H$\beta$ farther out in the BLR. Therefore, when the continuum changes, the response of the core of the line is more pronounced, and produces this effect (e.g., when the continuum strength increases the core of the line responds more strongly than the wings and the line appears narrower). \cite{barth15} and \cite{park12} have found that breathing, previously thought to appear over dynamical timescales of the BLR \citep[of order several years, see][]{obrien95, korista04, cackett06, bentz07}, can appear on timescales of the order of days to weeks following changes in the continuum. Although \cite{barth15} and \cite{park12} found that the broad H$\beta$ lines in most of the 15 objects they studied exhibited breathing behavior, it is important to note that ``anti-breathing" has also been observed: a study by \cite{guo14} found that the majority of their sample of 60 objects exhibited ``anti-breathing" behavior. \cite{zhang15} also examined an object in more depth that exhibits anti-breathing behavior in some of the broad emission lines, and \cite{wang20} find that different broad emission lines may show breathing or anti-breathing in the same object. The centroid velocity shift of these lines has been found to vary on similar timescales as breathing \cite{barth15}. This effect has been termed ``jitter", or ``radial velocity jitter". However, velocity jitter has also been observed to vary over longer timescales \citep[see][]{sergeev07}. In some cases jitter has been correlated with changes in the ionizing continuum or luminosity of the central engine, which may indicate that it may be a result of the BLR reverberation.

The role of radial velocity jitter is also vital in spectroscopic searches for binary SMBHs \citep[e.g.,][]{popovic12,dotti12,bogdanovic15}. Under the assumption that one SMBH in the binary is active and has a BLR, the broad emission lines will oscillate in velocity over an orbital period and can reveal sub-parsec binaries that cannot otherwise be telescopically resolved. The expected orbital periods of the binary SMBHs that can be found using radial velocity variations are on the order of several decades to centuries \citep[e.g.,][]{eracleous12}. Most of the promising spectroscopic times series currently available span under 30 years and are sparsely sampled \citep[][]{doan20, mohammed25}. Radial velocity jitter that is a result of the reverberation of the BLR can create apparent variations that can be mistaken for the signature of orbital motion. Therefore, radial velocity jitter of the emission line profiles must be characterized and understood in order to appropriately deal with this effect in SMBH binary searches. 

%Another point of interest is that the changes in the emission line profiles on the dynamical time, combined with a synthetic light curve (described more in Section \ref{subsec:lightmodel} and \ref{subsec:light}) may reveal a characteristic time in which the emission line profiles change. This characteristic time can then be compared with the dynamical time of the BLR or the light-crossing time of the BLR, and can reveal either BLR structural information or ionizing continuum information, respectively.

In this paper we present the design of a software pipeline that can simulate reverberation of a disk-like BLR, which is achieved by shining a model for the light curve from an AGN central engine on the BLR, and then calculating the subsequent time-dependent emission line profiles. This pipeline takes advantage of published physical models to specify the geometry, kinematics, emissivity laws, and other features of the BLR, and incorporates special and general relativistic effects, which capture asymmetries in the line profiles even when the BLR is axisymmetric. We focus specifically on a disk-like BLR in this work because this model is well developed theoretically and has substantial observational support. We justify the use of this model further in Section \ref{sec:model}.  Previous studies \citep[e.g.,][]{welsh91, robinson95} have not assumed a single model of the BLR, but instead used more general reverberation mapping techniques (such as transfer functions) to get better constraints on the geometry and kinematics using multiple proposed models. There are also a number of forward-modeling codes that are very sophisticated and can adopt a number of geometries and velocity fields in order to produce velocity response functions for a variety of applications; see \cite{pancoast11}, \cite{li13}, and \cite{rosborough24}. Such forward-modeling codes focus on obtaining the mass of the SMBH from the constraints on the geometry of BLR, and seek not to use the virial factor, $f$, that is used in traditional reverberation mapping. 

Our project and code are specifically targeting the disk-like geometry of the BLR and explore how the broad emission line profiles may change as a result of this geometry when it reverberates from the central ionizing continuum. Our code uses analytic prescriptions instead of photoionization grids to save computational time and incorporates the effects of special and general relativity in the calculation of the line profiles, which were shown to be important \citep{tremaine14}. Thus, our pipeline allows us to quickly explore a very large parameter space for the model. A more thorough exploration of a specific range of parameter combinations may have to employ more sophisticated codes like those cited above, after our code has identified that parameter space. We aim to use this pipeline in forthcoming papers to study jitter and breathing of the broad Balmer lines. Our broader goals are to exploit breathing and jitter to develop tests for the disk-like BLR model and to explore the possibility of developing a viable jitter prescription for other applications. The pipeline was developed and tested for one particular BLR geometry, but is modular and can be adapted to apply to other geometries and velocity fields of the BLR as needed.

This paper is the first in a series and describes the development of the machinery for this project--future papers will discuss science results. In Section \ref{sec:model}, we introduce the model we are using, as well as the important assumptions and features that are included within the model. In Section \ref{sec:overview}, we give a brief overview of the main modules of the pipeline and the process necessary to create the simulated emission line profiles. In Section \ref{sec:modules}, we detail the exact function of each module. In Section \ref{sec:conclusion}, we discuss the current stage of the pipeline and describe end-to-end tests. We close with a summary and a brief outline of future work at the end of Section \ref{sec:conclusion}.

%% The "ht!" tells LaTeX to put the figure "here" first, at the "top" next
%% and to override the normal way of calculating a float position.
%% The asterisk after "figure" tells the compiler to span multiple columns
%% if a two column style is selected.
\section{Description of the Model} \label{sec:model}
 
\subsection{Geometry and Structure of the Broad-Line Region} \label{subsec:geometry}
Our goal in this version of the pipeline is to simulate the observational signature of a disk-like BLR and compute the variability of the moments of the resulting line profiles. We are assuming that the BLR has the form of a Keplerian disk that is just an extension of the hot inner accretion disk that produces the ionizing continuum. In this pipeline, we make no assumptions regarding the hot inner accretion disk. The hot inner accretion disk can be described by any model that allows it to emit ionizing radiation that can reach the BLR either through direct or indirect illumination. We incorporate analytic prescriptions of physical effects, such as radiative transfer in the base of the disk wind, relativistic effects, and emissivities intrinsic to the BLR to understand the changes on the line profiles. 

A disk-like BLR is a well-developed physical model and is consistent with the observed behaviors noted in the previous section---thus this pipeline can be used to test the model further. The BLR in this model is a continuous medium that is consistent with a disk and/or wind, and reverberation mapping campaigns often show the signature of a reverberating disk \cite[e.g.,][]{horne21, bentz23}. The VLT Interferometer has also resolved the BLR in 3C 273 and revealed that the BLR looks like a disk \cite[e.g.,][]{gravity20}. Several authors have considered models for a disk-like BLR, with some of the earlier models coming from \cite{emmering92}, the series by \cite{collin-souffrin89}, \cite{murray95}, and \cite{elitzur14}. More modern and sophisticated models are based on the same idea and include those of \cite{waters16}, \cite{matthews20}, \cite{baskin18}, and \cite{naddaf22}. The disk-like model for this project combines and utilizes the ideas from the above papers. It is possible for  users of this pipeline to adopt other disk geometries; in such a case the user would have to prescribe the time sequences of emissivity maps accordingly through the apparent emissivity generator and brightness map modules (discussed in Section \ref{subsec:emissivity}).

In this work, we are making the following important assumptions:
\begin{itemize}
  \item The BLR is a flat disk centered on the continuum source.
  \item The continuum originates in a central source much smaller than the BLR---i.e., a point source \cite[][]{dumont90a}.
  \item The BLR is in the equatorial plane and the line-emitting region makes up the ``skin" (i.e., the surface emission layer) of the outer disk. The emission layer may be accelerating in the vertical direction to form a wind.
  \item The emission lines respond linearly (to first order) to variations of the ionizing continuum (see Section\ \ref{subsec:discussion} for further discussion). 
  \item The observed optical/ultra-violet (UV) continuum is a good proxy of the ionizing continuum.
  \item The light-travel time across the BLR is the most important time scale---the local gas response is effectively instantaneous with a response time of $\sim 10^{-10}$ seconds \citep[][]{peterson93}, and the light-travel time is much shorter than the dynamical time scale for the BLR.
  \item The structure and emission properties of the BLR do not change substantially on the time scale of the ionizing continuum variations.
\end{itemize}

\subsection{Calculation of the Line Profiles} \label{subsec:linecalc}
We follow the approach described in \cite{flohic12} to calculate the broad emission line profiles from a disk-like BLR. This formalism, represented in Eqn. \ref{flux density}, was obtained by combining the methods described in \cite{chen89} and \cite{chen89b} with those of \cite{murray95}. We adopt the geometry and coordinate system in Figure 1 of \cite{chen89} and assume that the line emitting region is an annulus defined by an inner and an outer radius. We adopt a convention in which the unprimed coordinates ($\xi, \phi$) refer to the frame of the disk. We numerically calculate the flux density of the line profile as seen by an observer at infinity through
\begin{equation}
    f_{\nu} \propto \int \int I_{\nu_e} \left(\xi, \phi, \nu \right) D^3(\xi, \phi) \Psi(\xi, \phi)
    d\phi \xi d\xi
    \label{flux density}
\end{equation}
where $\xi$ is the radial distance from the central massive black hole in gravitational radii ($\xi = r/r_g$, where $r_g = GM_{BH}/c^2$) and $\phi$ is the azimuthal angle in the disk. The specific intensity of the line has two components---the disk emissivity function $\epsilon(\xi, \phi)$ and the local line profile, which is assumed to be Gaussian. Thus, we write the specific intensity as
\begin{equation}
    I_{\nu_e} \propto \epsilon(\xi, \phi) \exp\left[- \frac{(\nu_e-\nu_0)^2}{2\sigma^2} \right]
\end{equation}
where $\sigma$ is the line broadening parameter, $\nu_e$ is the emitted frequency of the line photon, and $\nu_0$ is the rest frequency of the line photon.
In Eqn. \ref{flux density}, the Doppler factor, $D(\xi, \phi)$, defined as $D \equiv \nu/\nu_e$, depends on the velocities within the disk and the potential, and therefore takes into account the assumed Keplerian motion of the disk. The term $\Psi(\xi, \phi)$ in Eqn. \ref{flux density} takes into account light bending, whose effects on the line profile are very small for the range of parameters we are exploring (i.e., inclination angles $i$ less than 80$^{\circ}$ and emission from outside $\sim100r_g$). It is important to emphasize the physical effects that the model includes, namely the special and general relativistic time dilation, Keplerian motion of a disk-like BLR, and light bending. The inclusion of such effects leads to an extended red wing on the profiles, which enhances the redshift of the centroid velocity. The profile asymmetries are further amplified by beaming, which is also included in the profile calculation.

\subsection{Emissivity Law(s)} \label{subsec:emissivity}

We take the emissivity of the BLR to consist of a combination of multiplicative components:
\begin{equation}
    \epsilon(\xi, \phi) = \epsilon_0\; \epsilon_a(\xi)\;\epsilon_p(\xi, \phi)\;\beta(\xi, \phi)
\end{equation}
where $\epsilon_0$ is a scale factor, $\epsilon_a$ is the axisymmetric component set by the illumination from the central source, $\epsilon_p$ is the emissivity from non-axisymmetric perturbations within the disk (e.g., spiral arms or bright spots), and $\beta$ is the directional escape probability of photons originating in the disk.

\begin{figure}[ht!]
\epsscale{0.75}
\includegraphics[width=0.5\textwidth]{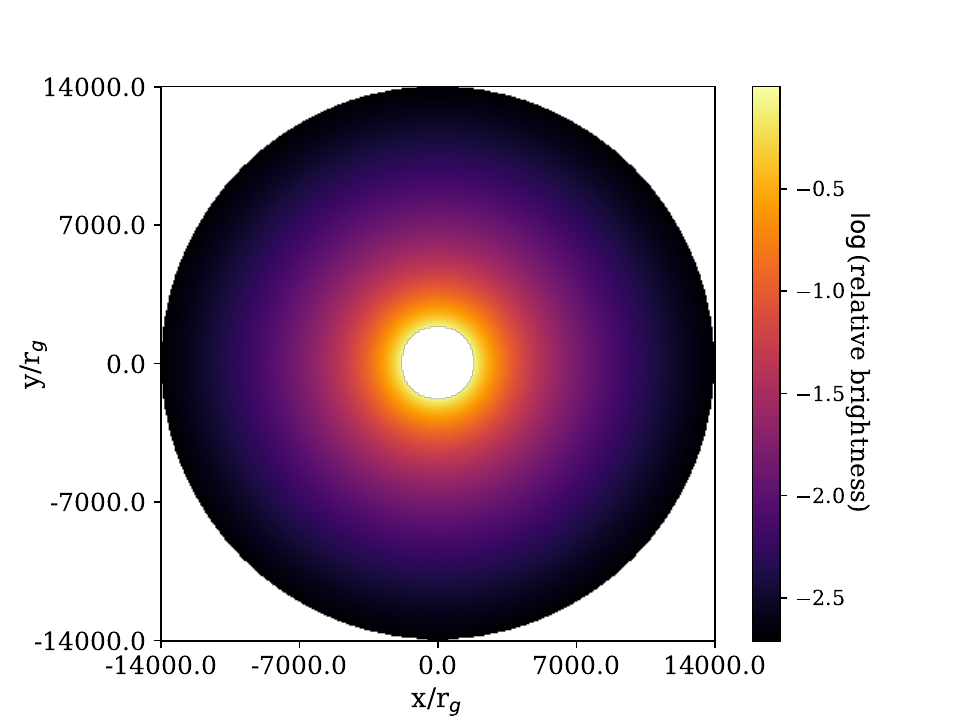}
\caption{Example of an emissivity map with a single power-law emissivity. The power-law goes as $r^{-3}$. The inner and outer radii of the BLR are given in units of the gravitational radius, as defined in Section \ref{subsec:linecalc}, and are 1750 and 14000 respectively. The brightness is relative to the flux at the inner radius.      
\label{fig:powerlaw}}
\end{figure}

\cite{collin-souffrin89} introduced both single and broken power-law approximations for $\epsilon_a$ in order to describe the results of their photoionization calculations. The power law is not a direct prediction of their physical model but a convenient parametrization for their numerical calculations, and \cite{dumont90b} have shown that the emitted flux of the H$\alpha$ line drops as $r^{-3}$ as a result of the geometric dilution of illumination. The use of parametric emissivity laws allows users of the pipeline to choose the line emissivity according to their goals, whether that be from numerical or analytic calculations, or empirical results.
Several papers, such as \cite{eracleous94}, \cite{eracleous03}, \cite{storchi-bergmann17}, and \cite{ward24}, have found that a power-law emissivity leads to good agreement between models and observed double-peaked line profiles. These studies have also found a similar agreement with the implementation of a broken power-law. Thus, the emissivity can be described by:
\begin{equation}
    \epsilon_a(\xi) = \left (\frac{\xi}{\xi_{in}}\right )^{-q}
\end{equation} 
for a radial power-law with index $q$, which includes a normalization scheme such that $\epsilon_a=1$ at $\xi=\xi_{in}$, and 
\begin{equation}
    \epsilon_a(\xi) = \begin{cases}
        \xi^{-q_1},  \xi_1<\xi<\xi_q\\
        \xi_q^{-(q_1-q_2)}\xi^{-q_2},  \xi_q<\xi<\xi_2
    \end{cases}
\end{equation}
for a broken power-law, where $\xi_q$ is the break radius, and $q_1$ and $q_2$ are the indices for the emissivity in their respective radial regions. Figure\ \ref{fig:powerlaw} shows an example of a disk with an emissivity of the form $\epsilon_a\propto r^{-3}$.
A separate, empirical law is inspired by reverberation mapping campaigns and can be parameterized as a bell-shaped function that peaks at some radius and declines in either direction \citep[e.g.,][]{horne21}. This law can be represented as 
\begin{equation}
    \epsilon_a(\xi) = \frac{1}{\sqrt{2\pi}\sigma_c}\exp\left[\frac{-(\xi-\xi_c)^2}{2\sigma_c^2} \right]
\end{equation}
where $\xi_c$ is the radial peak of the Gaussian, and $\sigma_c = \xi_c/2$. This bell-shaped function can subsequently be combined with a radial or broken power-law \citep[see][for examples]{sottocorno25} in order to reflect the empirical relation between the BLR characteristic radius and central engine luminosity, as discussed in Section \ref{sec:intro}.

Non-axisymmetric perturbations can be added to the disk, which can explain the shapes and variability of Balmer lines. Spiral arms can form in the disk as a result of gravitational instabilities within a self-gravitating disk or external perturbations \citep[e.g.,][]{adams89, livio91, chakrabarti93a}.

\begin{figure}[ht!]
\epsscale{0.75}
\includegraphics[width=0.5\textwidth]{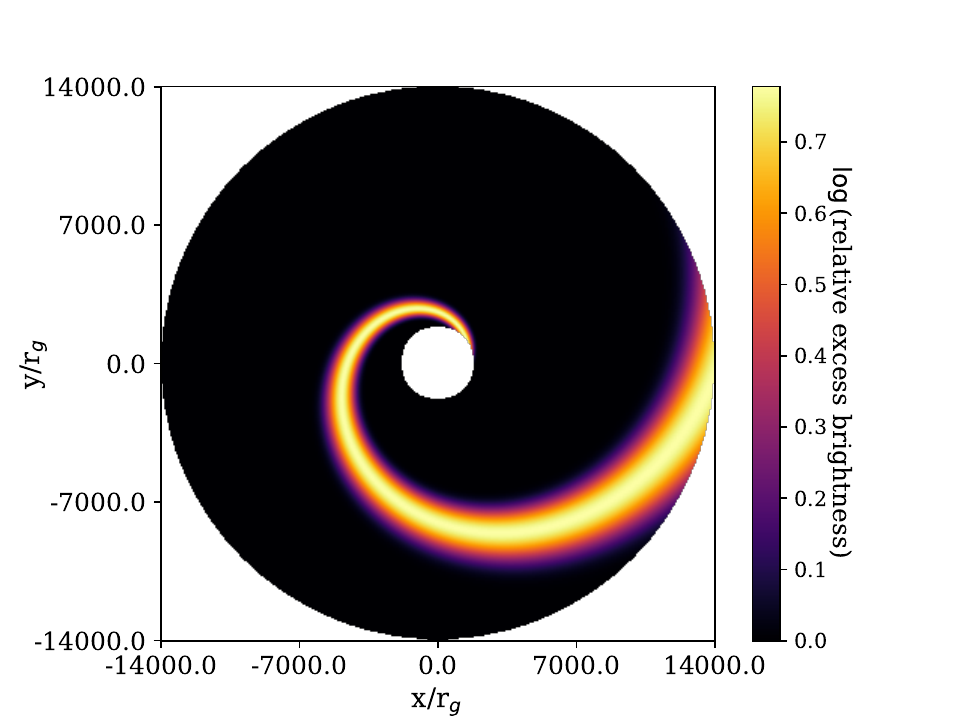}
\caption{Example of an emissivity map with a single-arm spiral. The same geometrical parameters from Figure\ \ref{fig:powerlaw} are applied here. Parameters of the spiral arm are: the azimuthal angle $\phi_0 = 30^{\circ}$, the pitch angle $p = 20^{\circ}$, the width $\delta = 30^{\circ}$, and the contrast $A=10.0$. The shading represents the excess brightness relative to the underlying BLR emissivity.    
\label{fig:spiralarm}}
\end{figure}

Previous works have found that the inclusion of a static or precessing spiral arm or bright spot on the surface of the disk leads to good fits of broad emission line profiles for single objects and samples of objects, \citep[e.g.,][]{storchi-bergmann03, gezari07, lewis10, storchi-bergmann17, ward24, ward25, rigamonti25}. Thus we use the spiral arm prescription of \cite{gilbert99} and \cite{storchi-bergmann03}:
\begin{equation}
\begin{split}
\epsilon_p(\xi, \phi) = \biggl\{1 +\frac{A}{2}\exp\biggl[-\frac{4\ln2}{\delta^2}(\phi-\psi_0)^2 \biggr] \\ 
    +\frac{A}{2}\exp\biggl[-\frac{4\ln2}{\delta^2}(2\pi-\phi+\psi_0)^2 \biggr]\biggr\}
\end{split}
\end{equation}
where $A$ describes the brightness contrast of the spiral arm relative to the underlying, axisymmetric disk, $\delta$ sets the azimuthal width of the spiral arm, and the exponentials describe the decay of the brightness of the spiral arm as a function of azimuthal distance from its ``ridge line" $\phi-\psi_0$. The angle $\psi_0$ describes the ridge line of the spiral arm through
\begin{equation}
    \psi_0 = \phi_0+\frac{\ln(\xi/\xi_{sp})}{\tan p }
\end{equation}
where $\xi_{sp}$ is a fiducial radius (typically the inner radius of the disk), $p$ is the pitch angle of the spiral arm, and $\phi_o$ sets the orientation of the pattern. 
An example of a spiral arm emissivity is given in Figure\ \ref{fig:spiralarm}. The values of the spiral arm parameters are constrained by fits to observed line profiles \citep[see][references therein]{schimoia12, schimoia17, storchi-bergmann17, ward24, ward25}. We have narrowed the parameter range to $0<A<8$, $-30<p<90$, $50<\phi_0<300$, and $20<\delta<90$ according to these studies. 

The final component of the overall emissivity of the disk is the directional escape probability. In the models of \cite{murray95}, the broad lines are produced at the base of a disk wind and must travel through the accelerating emission layer to escape. We use the following expression for the escape probability \citep{hamann93, murray98}
\begin{equation}
    \beta(\tau_{\nu_e}) = (1-e^{-\tau_{\nu_e}})/\tau_{\nu_e}
\end{equation}

\begin{figure}[ht!]
\epsscale{0.75}
\includegraphics[width=0.5\textwidth]{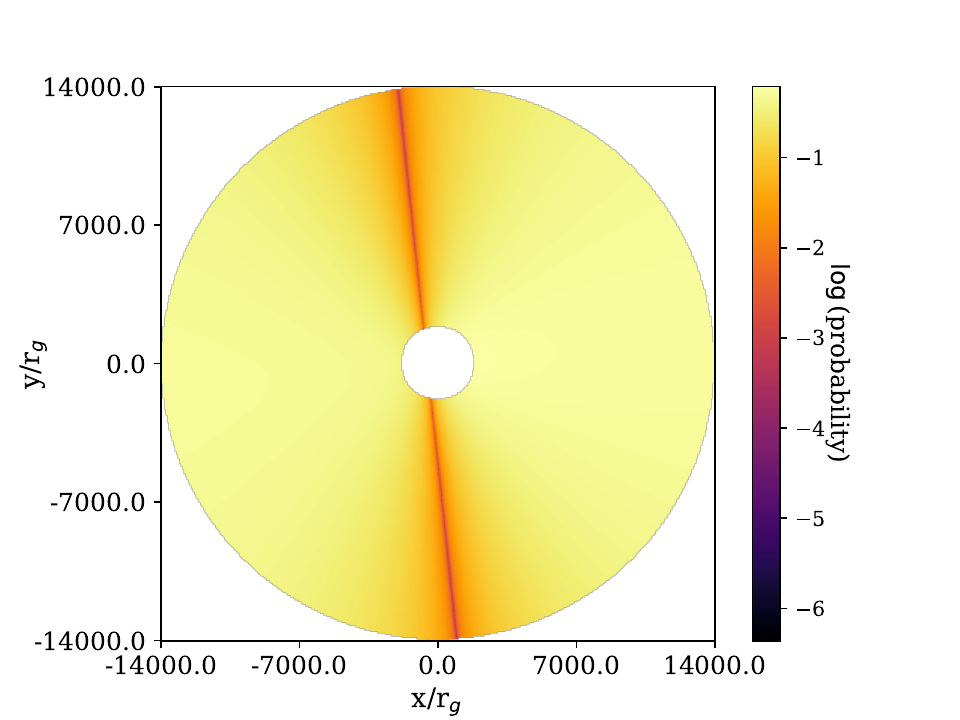}
\caption{Example of the escape probability map of line photons emitted at a specific location on the disk. The same geometrical parameters from Figure\ \ref{fig:powerlaw} are applied here. The radiative transfer parameters from equation (\ref{tau}) are $\tau_0 = 10.0$ and $\eta = 0.5$, as constrained by the work of \cite{chajet13}. The observer is to the right of the disk, with an inclination angle of 30$^{\circ}$ between the direction of the observer and the axis of the disk.    
\label{fig:radtransfer}}
\end{figure}

In order to calculate the position-dependent optical depth $\tau_{\nu_e}$, we utilize the following expression from \cite{flohic12} \citep[see also][]{murray98, chajet13}
\begin{equation}
    \tau(\xi, \phi) = \frac{\tau_0}{Q_0} \left( \frac{\xi}{1000} \right)^{3/2-\eta}
    \label{tau}
\end{equation}
where $\tau_0$ is the total optical depth of the surface emission layer perpendicular to the disk plane at $\xi = 1000$, $\eta$ describes the dependence of density in the emission layer on radius, and $Q_0$ represents the analytic prescription of the projected strain tensor of a disk wind from \cite{murray95} \citep[see Eqn. 8 in][for the full expression for $Q_0$ and its dependencies on the inclination angle between the observer and the axis of the disk, $i$]{flohic12}. An illustrative example of how the escape probability affects the emissivity of the disk can be seen in Figure\ \ref{fig:radtransfer}.

Through the generation of these maps, we dictate how the illumination sets the emissivity of the BLR. Users can modify the relevant modules to produce maps that encode geometric or emissivity information according to their own goals. For example, one can adopt an non-axisymmetric illuminaiton pattern for the BLR and compute the maps accordingly.

\subsection{Models for the Light Curve} \label{subsec:lightmodel}
We utilize light curve templates or synthetic light curves from statistical models in order to simulate reverberation of the BLR. The origin and uses of light curve templates is further discussed in Section \ref{subsec:light}. For the purposes of creating synthetic light curves, we utilize the statistical and empirical Damped Random Walk (DRW) model for light curves of AGNs \citep[e.g.,][]{kelly09, macleod12}. \cite{macleod10} and \cite{suberlak21} show that the model parameters of the DRW can be linked to physical properties of the central engine such as the black hole mass and luminosity of the AGN, and more details are given in Section \ref{subsec:light}. The use of the connections found by \cite{macleod10} and \cite{suberlak21} is vital in ensuring that the synthetic light curves created also remain physically realistic. Examples of light curves are shown in Section \ref{subsec:light}.
The use of synthetic light curves allows us to create long, uninterrupted time series of output parameters and to explore the behavior of the models over a wide range of input parameters. 

\section{Overview of the Pipeline} \label{sec:overview}

\begin{figure*}[ht!]
\includegraphics[width=\textwidth]{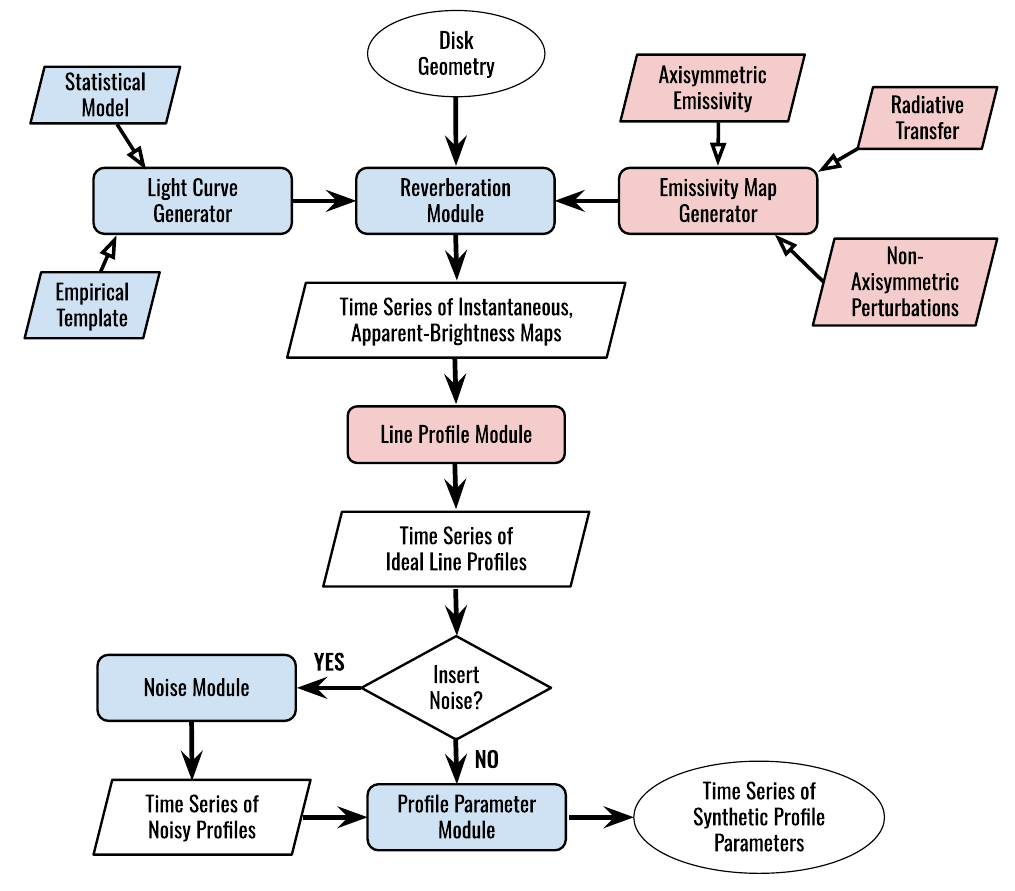}
\caption{The basic process of the pipeline, with arrows to indicate the progression. Boxes in blue are represent statistical or empirical processes, boxes in red represent physical processes, and boxes in white represent outputs that can be utilized and examined.
\label{fig:pipeline}}
\end{figure*}

The pipeline assumes a specific geometry and specific emission properties for a disk-like BLR and takes in a light curve for the illuminating source. It then creates a time sequence of line profiles and subsequently measures model-independent shape parameters of these profiles. The process is illustrated in Figure\ \ref{fig:pipeline}.
The light curve generator, discussed in Section \ref{subsec:light}, allows the user to choose a synthetic or empirical light curve template, and the emissivity modules discussed in Section \ref{subsec:emissivity} allow a choice of components for the final emissivity law. The reverberation module, discussed in Section \ref{subsec:reverb}, uses the adopted disk geometry and generates instantaneous {\it apparent} illumination maps using the corresponding geometrical time delay of the light as it crosses the BLR. The line profile module of Section \ref{subsec:line} uses these brightness maps, as well as the disk geometry, and generates a time series of smooth emission line profiles. The profile parameter module in Section \ref{subsec:analysis} measures shape parameters of each emission line profile. There is an option to add simulated noise to the smooth emission line profiles, which is discussed in Section \ref{subsec:noise}. The end result is a time series of synthetic profiles and profile parameters that allows us to quantify how these broad emission lines change over time, and how they may `jitter' or `breathe'.

\section{Modules of the Pipeline} \label{sec:modules}

\subsection{Light Curves} \label{subsec:light}
Light curves are utilized in creating the reverberation maps of the disk. There are two types of light curves that can serve as inputs to the pipeline --- synthetic light curves and empirical templates. Both types undergo processing in the Light Curve Generator Module from Figure\ \ref{fig:pipeline} by normalizing by the mean, which emphasizes fluctuations in local illumination relative to the mean of the light curve. Furthermore, this makes the use of multiple input light curves from different sources comparable to each other as it emphasizes the variabilities and fluctuations.

As discussed in Section \ref{subsec:lightmodel}, a synthetic light curve is created using the statistical and empirical DRW model, and is part of the Statistical Model portion of Figure\ \ref{fig:pipeline}. However, in the future, other models for AGN light curves may also be utilized in this module. There are three input parameters for the DRW model: the structure function at infinity $SF_{\infty}$, a characteristic timescale $\tau$, and a mean absolute magnitude in the i-band $M_i$, as described in \cite{kelly09} and \cite{macleod10}. This model successfully recreates the photometric variability of AGNs and nearby quasars. \cite{macleod10} and \cite{suberlak21} found that there is no dependence of the input parameters on redshift, but there are empirical relations between the rest-frame wavelength, black hole mass, and absolute mean-magnitude in the i-band. The damped random walk parameters can be written as
\begin{equation}
\begin{split}
    \log f = A + B\log\left(\frac{\lambda_{RF}}{4000\;{\rm  \AA}}\right) + C(M_i+23)\\ + D\log\left(\frac{M_{BH}}{10^9\;{\rm M}_{\odot}}\right),
\end{split}
\end{equation}
where $f$ represents the parameter of interest, $SF_{\infty}$ or $\tau$. The coefficients $A$, $B$, $C$, $D$ take different values depending on which damped random walk parameter we wish to compute. We utilize the prescriptions in \cite{kelly09} for implementing this auto-regressive process. 

\begin{figure}[ht!]
\epsscale{0.75}
\includegraphics[width=0.5\textwidth]{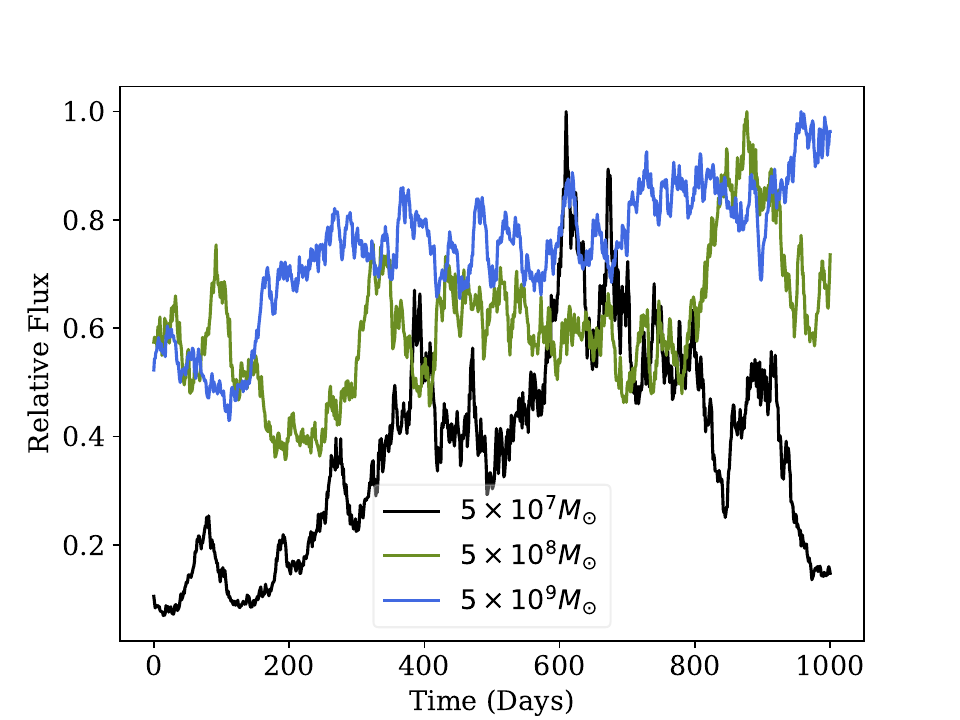}
\caption{Examples of light curves generated using the Damped Random Walk model, as described in \cite{macleod10}. The input parameters were a rest-frame wavelength of 900 \AA \ and an Eddington ratio of 0.1. We varied the mass of the central black hole in the range of $10^7-10^9 M_{\odot}$ and normalized the light curves by their maximum value.
\label{fig:synlight}}
\end{figure}

In order to get an absolute i-band magnitude, we calculate the bolometric luminosity of the AGN under the assumption that its black hole is accreting with a chosen Eddington ratio, $L/L_{Edd}$. We use a correction factor described in \cite{richards06} to get the i-band luminosity, with a median wavelength of 7625 \AA\ for the i-band. We chose a rest-frame wavelength of approximately 912 \AA\ to model the ionizing continuum that is primarily responsible for the line emission. This is vital in ensuring that the light curve generated refers to the energy band just above 1 Ry. Examples of light curves with varying input parameters can be seen in Figure\ \ref{fig:synlight}. Testing of the parameter space of the model will utilize these general synthetic light curves.

\begin{figure}[ht!]
\epsscale{0.75}
\includegraphics[width=0.5\textwidth]{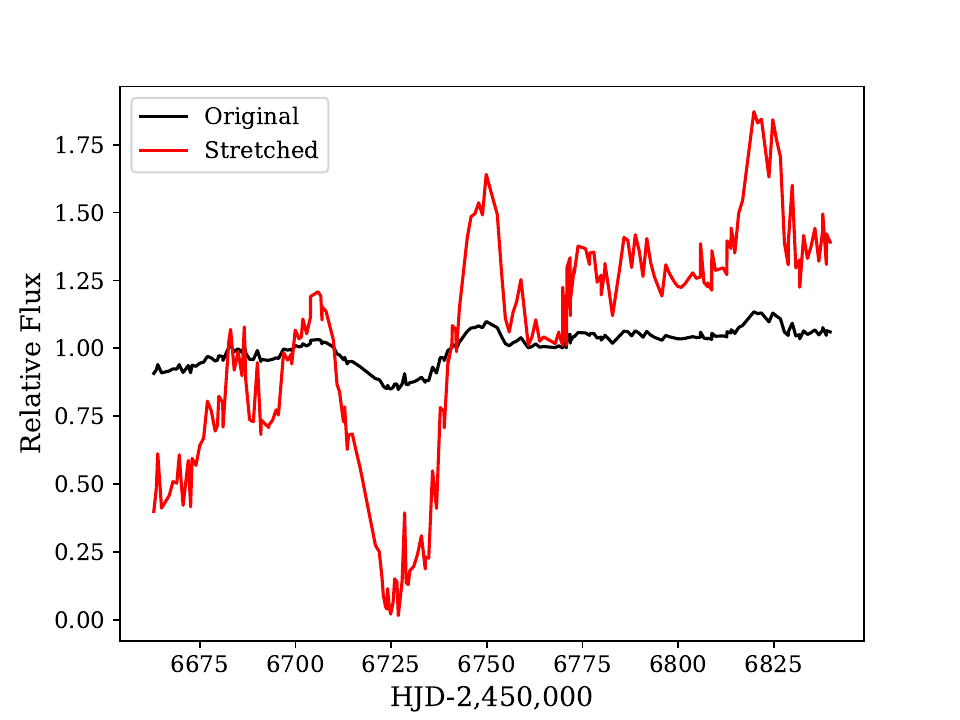}
\caption{Light curve of NGC 5548 measured at 5100 \AA, which was used as an input in some of the simulations. The original light curve is given in black and a modified version in which it is stretched is given in red and reflects the ``bluer when brighter" observations described in Section \ref{subsec:light}.  
\label{fig:ngc5548light}}
\end{figure}

Empirical light curve templates can also be used in the pipeline for testing specific scenarios for comparison to observed parameters. It is important to note that empirical templates undergo additional processing in the Light Curve Generator Module seen in Figure\ \ref{fig:pipeline}. The pipeline `stretches' the amplitude of the fluctuations as studies \citep[e.g.,][]{macleod12, sun15} have shown that at shorter wavelengths, the variability amplitude increases, a trend known commonly as `bluer when brighter'. Vertical stretching of the light curve, therefore, gives a better representation of the ionizing continuum, and produces the same shape in the light curve but with more pronounced fluctuations, as demonstrated in Figure\ \ref{fig:ngc5548light}. 

An empirical light curve used for testing against observations was the measured continuum of NGC 5548 obtained in an intensive reverberation mapping campaign, as detailed in \cite{kriss19} and shown in Figure\ \ref{fig:ngc5548light}. 
The light curve was produced from flux observations at 5100 \AA\ and discussed in \cite{derosa18}.

\subsection{Reverberation} \label{subsec:reverb}
Under the assumption that the BLR is a flat disk centered on the continuum source, we use the formula from \cite{peterson93} for the time delay:
\begin{equation}
    \Delta t = \frac{r}{c}(1-\sin i\cos \phi)
    \label{time}
\end{equation}
where $r$ is the radius in the BLR, $i$ is the inclination angle, and $\phi$ denotes the azimuthal angle in the disk.
This time delay captures the difference in path lengths of rays that come directly from the continuum source and rays that first go from the continuum source to a particular location in the disk, and then to the observer.

We calculate the inner and outer radii of the BLR based on physical considerations. Using the empirical relation described in \cite{bentz13}, the characteristic radius is related to the luminosity through
\begin{equation}
    \log(R_{BLR}) = K + \alpha \log\left[\frac{\lambda L_{\lambda}(5100\;{\rm \AA})}{10^{44}\;{\rm erg\;s^{-1}\;cm^{-2}}}\right]
    \label{BLRradius}
\end{equation}
where $K$ and $\alpha$ are the best fit values from the study and $\lambda L_{\lambda}$ is the luminosity measured at 5100~\AA. \cite{richards06} provide a conversion factor, which was $12.17$, to connect bolometric luminosity to the luminosity measured at 5100~\AA, which is used in equation (\ref{BLRradius}). The outer radius is set to be five times this value for the purposes of testing the pipeline, as studies such as \cite{koshida14}, \cite{yang20}, and \cite{lira24} have found that dust reverberation from the torus has time delays that are approximately five times longer than the BLR reverberation timescale. The inner radius of the BLR for the tests carried out here is set to be exactly 500~$r_g$, which is another empirical relation found in \cite{storchi-bergmann17}. The assumption we have made, especially the linear responsivity and monotonically decreasing emissivity, and the fixed values that we have adopted for the inner and outer radii are suitable only for testing the pipeline and do not allow us to self-consistently compute a ``characteristic radius'' of the BLR. We defer this exercise to future work.

\begin{figure}[ht!]
\epsscale{0.75}
\includegraphics[width=0.5\textwidth]{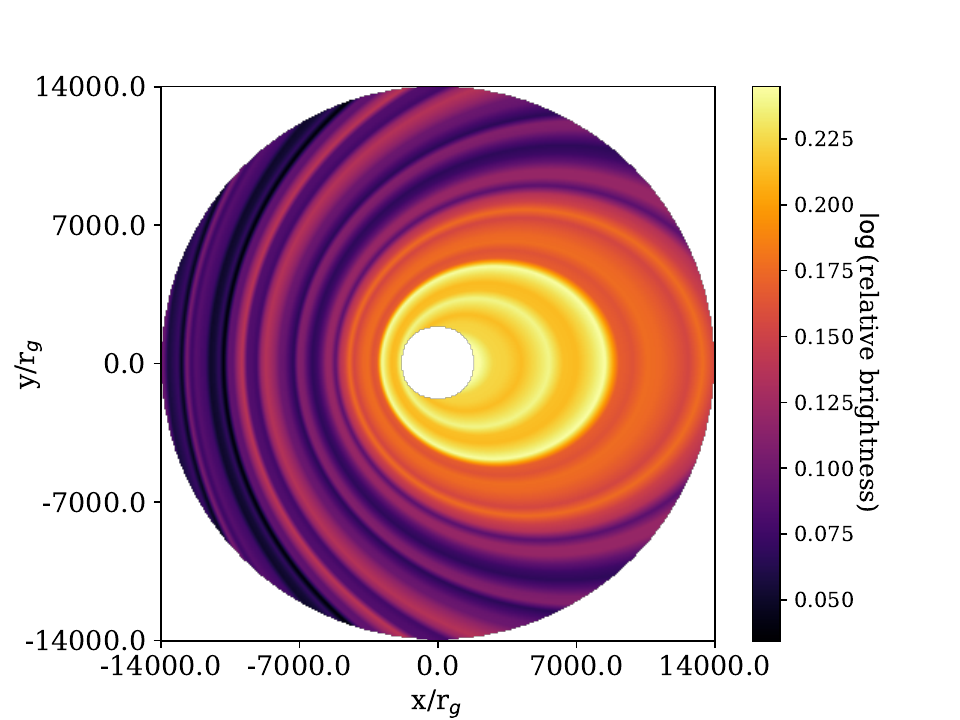}
\caption{Example of a reverberation map that utilizes time delays with a given light curve. The same geometrical parameters from Figure\ \ref{fig:powerlaw} are applied here. The shading on the figure represents the apparent brightness once we take into account the time it takes for a ray to leave the central source, hit a location on the disk, and for the subsequent reprocessing to reach the observer, who is to the right of the disk. This particular time delay map utilizes the DRW synthetic light curve depicted in Figure\ \ref{fig:synlight} in black, and shows fluctuations in local illumination. 
\label{fig:timedelay}}
\end{figure}

In the Reverberation module in Figure\ \ref{fig:pipeline}, we calculate a time delay for each location on the disk and connect that delay with a point on the light curve described in Section \ref{subsec:light}. The module assigns a brightness or instantaneous illumination for a particular location in the BLR by using the corresponding intensity of the light curve and the light travel time to the BLR pixel. The result is a reverberation map of the disk for one particular instant, with an example shown in Figure\ \ref{fig:timedelay}. This, and other reverberation maps, are proportional to the flux arriving at a specific BLR element. 

\begin{figure}[ht!]
\epsscale{0.75}
\includegraphics[width=0.5\textwidth]{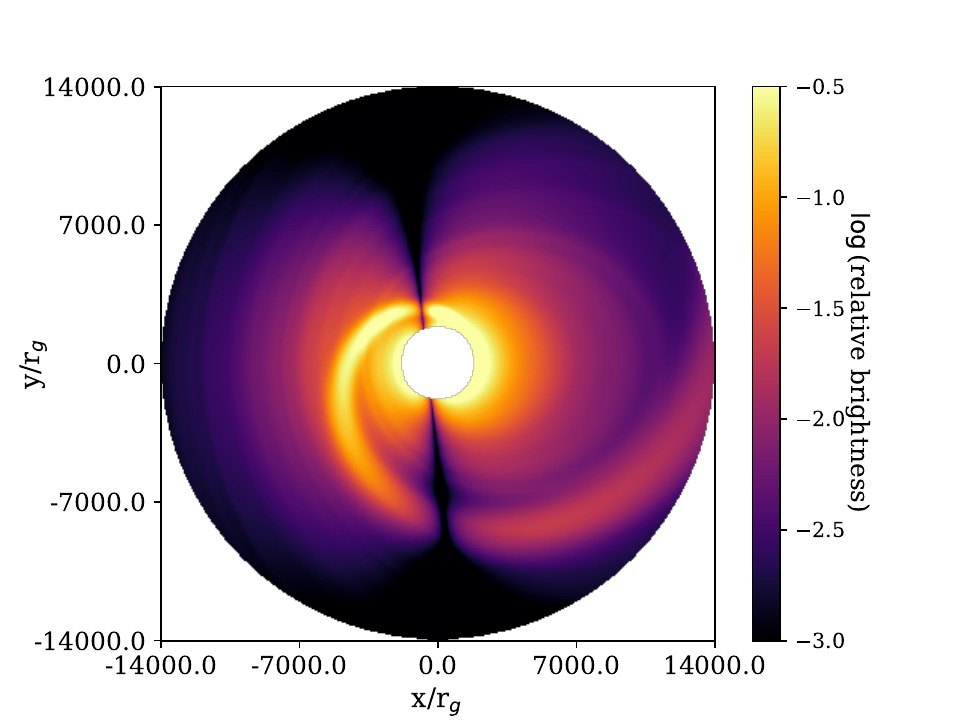}
\caption{Example of an instantaneous apparent brightness map. The same geometrical parameters from Figure\ \ref{fig:powerlaw} are applied here. This figure is the product of the maps of Figures\ \ref{fig:powerlaw}, \ref{fig:spiralarm}, \ref{fig:radtransfer}, and \ref{fig:timedelay}. The shading on the figure represents the instantaneous brightness of the disk relative to the flux from the axisymmetric component, $\epsilon_a$, at the inner radius, but the range has been adjusted to highlight the reverberation echoes.
\label{fig:brightness}}
\end{figure}

The reverberation map is multiplied by the apparent emissivity map discussed in Section \ref{subsec:emissivity} to produce the instantaneous apparent brightness map. An example of this final map is shown in Figure\ \ref{fig:brightness}. The instantaneous apparent brightness map is used in the calculation of the synthetic broad emission line profiles.

\subsection{Line Profiles} \label{subsec:line}
\begin{figure}[ht!]
\epsscale{0.75}
\includegraphics[width=0.5\textwidth]{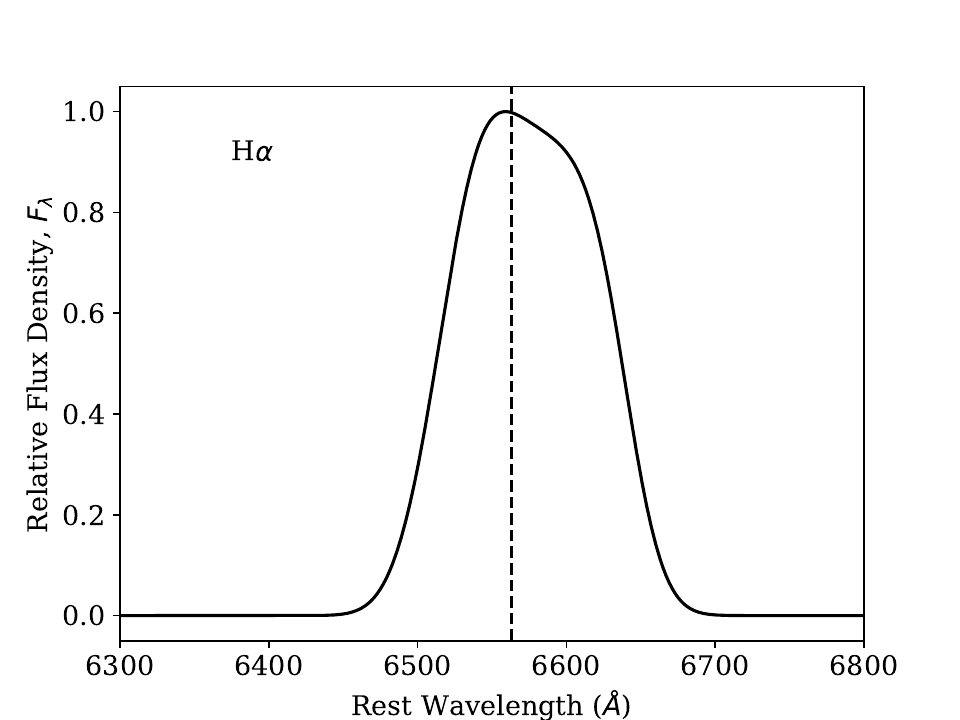}
\caption{Example of a smooth emission line profile, corresponding to a disk with an inclination angle of $30 ^{\circ}$ and the same geometrical parameters as described in Figure\ \ref{fig:powerlaw}. This profile comes from the instantaneous apparent brightness map of Figure\ \ref{fig:brightness}, with all of its underlying emissivities and time delays, and no adjustment to the range. The dashed vertical line shows the rest wavelength of H$\alpha$ at 6563~\AA.
\label{fig:emission}}
\end{figure}

The instantaneous apparent brightness maps are passed to the Line Profile Module seen in Figure\ \ref{fig:pipeline}, whose original numerical methods and model were described in Section \ref{subsec:linecalc}. Other input parameters include the inner and outer BLR radius, inclination angle, and rest wavelength of the emission line. The numerical integration grid uses logarithmic steps in radius to capture the rapid radial variation of emissivity and linear steps in azimuth. The output is a smooth, normalized emission line profile, as demonstrated in Figure\ \ref{fig:emission}. A separate profile is computed at each time step, corresponding to each instantaneous apparent brightness map. For the purposes of setting up and testing the pipeline, we have simulated the broad H$\alpha$ line, as seen in Figure\ \ref{fig:emission}.

\subsection{Profile Parameter Module} \label{subsec:analysis}
One can measure shape parameters from the emission line profiles in order to characterize them using the Profile Parameter Module in Figure\ \ref{fig:pipeline}. These parameters can be extracted before and after the addition of the noise in the profiles in order to examine the systematic distortions produced by noise (discussed in Section\ \ref{subsec:noise}). There are three parameters of interest, which are associated with the first three moments of the profile: the centroid velocity shift, the velocity dispersion, and the Pearson skewness coefficient. 

The first moment (centroid) of the profile is defined as
\begin{equation}
    \langle \lambda \rangle = K\sum_i \lambda_i f_i
\end{equation}
where $K$ is the normalization constant defined by 
\begin{equation}
    \frac{1}{K} = \sum_i f_i 
\end{equation}
and $\lambda_i$ and $f_i$ are discrete wavelengths and flux densities. By extracting this parameter and creating a time sequence, we can observe how the profile `jitters' over time. The first moment is converted to a velocity shift by:
\begin{equation}
    v_{cen} = c \frac{(\langle \lambda \rangle-\lambda_0)}{\lambda_0}
\end{equation}
where $v_{cen}$ is the resulting centroid velocity, $c$ is the speed of light, $\langle \lambda \rangle$ is the centroid wavelength for an emission profile, and $\lambda_0$ is the rest wavelength. 

The second moment is defined as
\begin{equation}
    \mu_2 = K \sum_i (\lambda_i-\langle \lambda \rangle)^2 f_i
\end{equation}
and is related to the standard deviation of the line profile via $\mu_2 = \sigma^2$. The velocity dispersion is then computed as: 
\begin{equation}
    v_{disp} = c \frac{\sigma}{\lambda_0}
\end{equation}

The Pearson skewness coefficient is defined as
\begin{equation}
p = \frac{(\langle \lambda \rangle -\lambda_m)}{\mu_2^{1/2}} 
\end{equation}
where $\lambda_m$ is the median wavelength of the profile. This parameter is a robust proxy of the third moment of the profile, and can therefore track its skewness.

The Profile Parameter Module is the last module in the pipeline, and the user is left with multiple outputs. As seen in Figure\ \ref{fig:pipeline}, there are the time series of instantaneous brightness maps, the time series of the ideal line profiles, the time series of the noisy profiles (if choosing to add noise), and the time series of synthetic profile parameters which can be further used for other tests. An illustrative example of these parameters and a discussion of a full test of the pipeline can be found in Section \ \ref{subsec:test}.  

\subsection{Synthetic Noise} \label{subsec:noise}
Adding noise to these profiles turns them into more faithful representations of actual observations of the emission lines from AGN. The addition of noise also reveals how much profile parameters can be {\it systematically} distorted. The user can choose if noise gets injected into the broad profiles, as seen by the decision tree in Figure\ \ref{fig:pipeline}. 

\begin{figure}[ht!]
\epsscale{0.75}
\includegraphics[width=0.5\textwidth]{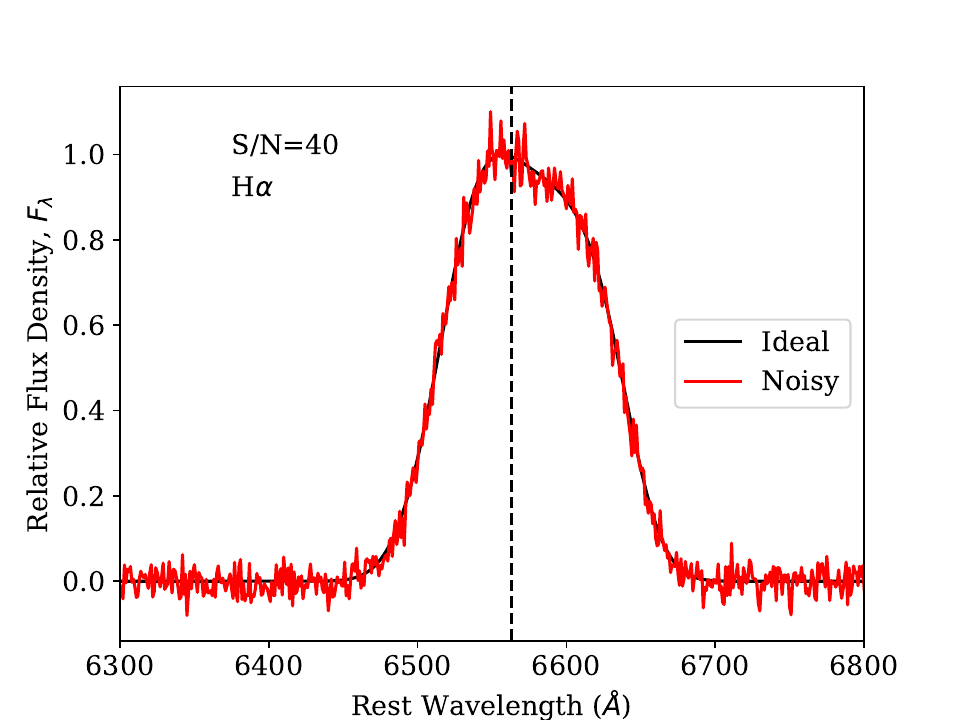}
\caption{The same emission line profile seen in Figure\ \ref{fig:emission}, now with noise overlaid in red after adopting a signal-to-noise ratio of 40 in the continuum. The adopted continuum level has the same flux density as the peak of the line. The dashed vertical line shows the rest wavelength of H$\alpha$ at 6563 \AA.
\label{fig:noisyemission}}
\end{figure}

To make the noise more realistic, we first add a power-law continuum of the form 
\begin{equation}
    f(\lambda) = b\left(\frac{\lambda}{\lambda_n}\right)^{\alpha}
\end{equation}
Here, $b$ is a normalization constant and $\lambda_n$ is the fiducial wavelength at which the user specifies the signal-to-noise ratio (S/N). The power-law index $\alpha$ can range from $-3$ to 3 \citep[][]{edelson86}. For the purposes of the experiments below, we chose $\alpha=-1.5$ and normalize at 6563 \AA. The relative strength of the continuum is an adjustable parameter and the S/N scales with the square root of the flux density to mimic Poisson noise. We obtain the noisy value through the use of a random normal distribution generator. This normal distribution has a mean equal the local flux at any given pixel in the spectrum, and a standard deviation calculated using the local value of the S/N. An example of a noisy emission line profile is shown in Figure\ \ref{fig:noisyemission}.

\begin{figure}[ht!]
\epsscale{0.75}
\includegraphics[width=0.5\textwidth]{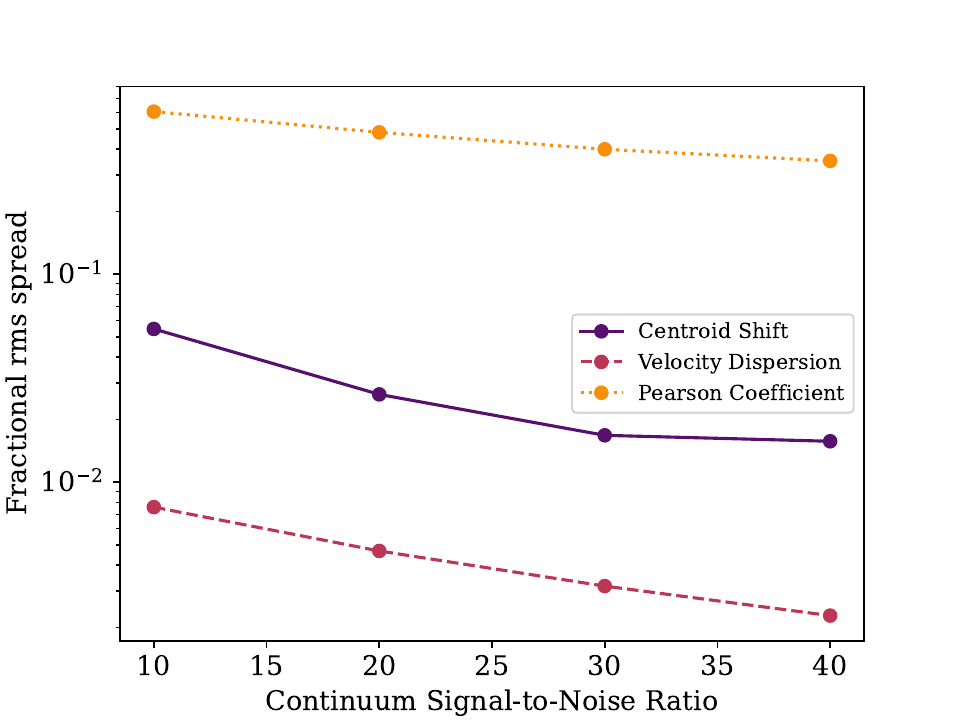}
\caption{The change in the fractional rms spread of the three parameters over a large number of realizations as the signal-to-noise ratio of the continuum increases. The solid purple line refers to the centroid velocity shift, the dashed pink line refers to the velocity dispersion, and the dotted orange line refers to the Pearson Skewness coefficient. 
\label{fig:sntest}}
\end{figure}

We evaluated the effects of S/N on the measured profile parameters. We created a large number of realizations of a noisy profile from one single ideal profile and measured the line profile moments from each realization. We then examined the distribution of the measured moments of the line profiles to quantify the fractional uncertainty resulting from noise. We repeated the exercise for S/N = $10, 20, 30, 40$, and found that the fractional rms decreased with increasing S/N, as illustrated in Figure\ \ref{fig:sntest}.

\begin{figure*}[ht!]
\centerline{\includegraphics[scale=0.7]{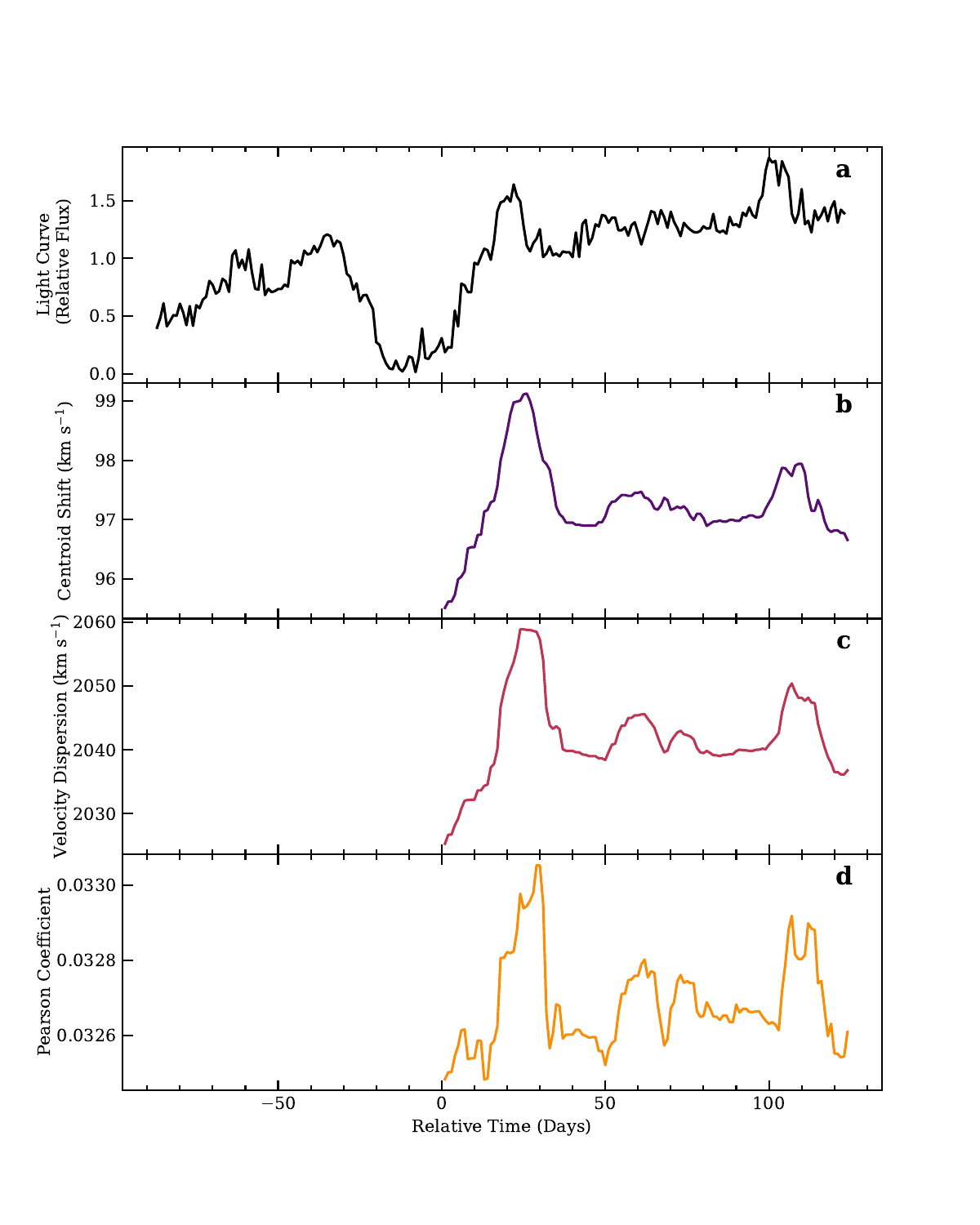}}
\caption{(a) The empirical light curve first introduced in Figure\ \ref{fig:ngc5548light}. It has also been through the preliminary processing and is normalized by its mean. The simulated time series begins at $t=0$, when every part of the disk has had a chance to reverberate. Points in the light curve at $t<0$ represent the full ``history" determining the first point of the output at $t=0$. (b) The centroid velocity shift as a function of time for the ideal line profiles. The error of the values is the 1$\sigma$ deviation of the difference between the ideal and noisy centroid velocity shifts, which was approximately 9 km s$^{-1}$, for the signal to noise value of 40 for the continuum. (c) The velocity dispersion as a function of time for the ideal line profiles. The error is the 1$\sigma$ deviation of the difference between the ideal and noisy velocity dispersions, which was approximately 4.6 km s$^{-1}$. (d) The Pearson skewness coefficient as a function of time for the ideal line profiles. The error is the 1$\sigma$ deviation of the difference between the ideal and noisy Pearson coefficients, which was approximately 0.003.
\label{fig:pipelinetest_sym}}
\end{figure*}

\begin{figure*}[ht!]
\centerline{\includegraphics[scale=0.7]{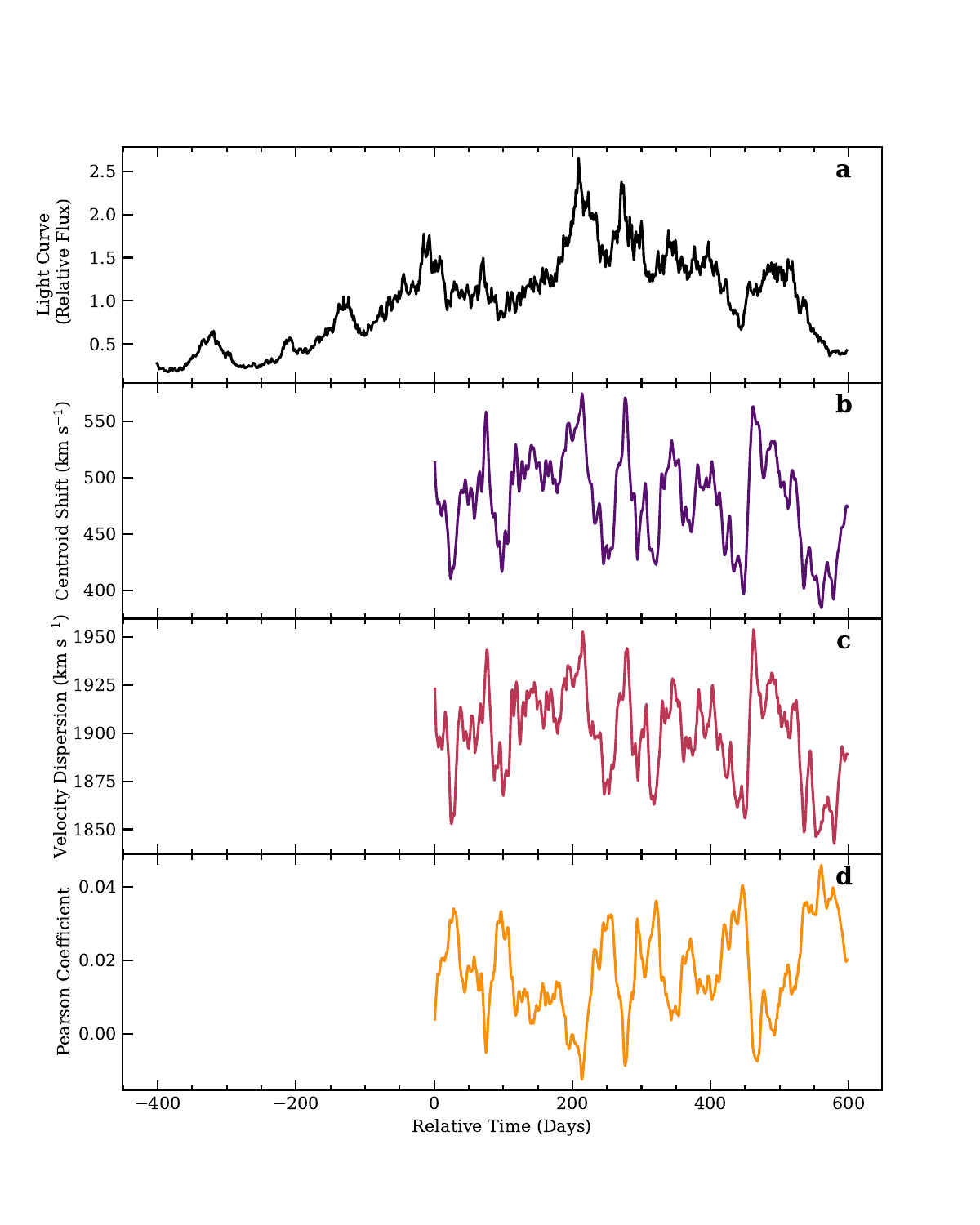}}
\caption{(a) Same as in Figure\ \ref{fig:pipelinetest_sym} for the synthetic DRW light curve first introduced in Figure\ \ref{fig:synlight} (black line). (b) The centroid velocity shift as a function of time for the ideal line profiles. The error value is the 1$\sigma$ deviation of the difference between the ideal and noisy centroid velocity shifts, which was approximately 9.8 km s$^{-1}$, for the signal to noise value of 40 for the continuum. (c) The velocity dispersion as a function of time for the ideal line profiles. The error is the 1$\sigma$ deviation of the difference between the ideal and noisy velocity dispersions, which was approximately 5.9 km s$^{-1}$. (d) The Pearson skewness coefficient as a function of time for the ideal line profiles. The error is the 1$\sigma$ deviation of the difference between the ideal and noisy Pearson coefficients, which was approximately 0.003.
\label{fig:pipelinetest}}
\end{figure*}

\section{Discussion and Conclusions} \label{sec:conclusion}
\subsection{Limitations of the Pipeline} \label{subsec:discussion}
The current limitations of the pipeline are set by the assumptions listed in Section \ref{sec:model}. Some of the limiting assumptions are that the disk is thin, and that the spiral arm is a brightness perturbation. The use of spiral arms is a departure from the assumption that the gas moves in a simple circular and Keplerian velocity field, and the departures from Keplerian rotation associated with the spiral arm are not included.  

We make a significant assumption regarding the responsivity of the disk, which describes how the line responds to changes in the strength of the ionizing continuum, i.e.,
\begin{equation}
    R = \frac{d \ln \epsilon_{line}}{d \ln f_{cont}}
\end{equation}
where $\epsilon_{line}$ is the line emissivity and $f_{cont}$ is the flux of the ionizing continuum at some fiducial wavelength impinging on the surface of the disk. We assume that the responsivity is linear everywhere, i.e., a given fractional change in the intensity of the ionizing continuum produces the same fractional change in the emissivity of the line at that location. Future work on this project will incorporate a non-linear responsivity of the disk using methods described in \cite{korista04} and \cite{goad14}. The use of non-linear responsivities would reflect more accurately the observations of the mean and rms spectra from reverberation mapping campaigns, as discussed in \cite{li25}.

\subsection{Tests of the Pipeline} \label{subsec:test}
We have carried out two end-to-end tests of the pipeline using both a synthetic light curve and an empirical template. Both tests have a S/N of 40 for the continuum, and the rms spread is calculated as described in Section \ref{subsec:noise}. More specific information regarding the values for each test can be found in the captions for Figures\ \ref{fig:pipelinetest_sym} and \ref{fig:pipelinetest}. The parameters used for these tests were chosen to showcase the capabilities of the pipeline; as such they are generally plausible but do not represent any specific system. The first test utilizes the light curve template presented in Figure\ \ref{fig:ngc5548light}, and has the same disk parameters as those presented in Figure\ \ref{fig:powerlaw}, which makes the disk axisymmetric. This test produces the time series of profile parameters presented in Figure\ \ref{fig:pipelinetest_sym}. Velocity jitter, as discussed in Section \ref{sec:intro} and seen in Figure\ \ref{fig:pipelinetest_sym}b, represents the fluctuations of the centroid of the line profile with time. We see that all three parameters rise sharply at the beginning in response to driving/ionizing light curve. Furthermore, we find that the line profile parameter time series seem to resemble the input driving light curve with only a small time delay-this reaffirms the process of our `reverberation' and may illustrate that the underlying emissivities and Keplerian velocity distribution are not the dominant process in the changes in the emission line profiles. Studies such as \cite{barth15} have found that velocity jitter can appear on BLR light-crossing timescales (days to weeks), and that the pattern of these changes seems to be correlated with the shape of the continuum light curve, which is presented in Figure\ \ref{fig:pipelinetest_sym}a. After a visual examination of the centroid velocity shifts shown in Figure\ \ref{fig:pipelinetest_sym}b, we confirm that the jitter has the same variation of the continuum light curve, which was seen in \cite{barth15}. Similarly, the amplitudes of the centroid velocity shift are of the same magnitude as those observed by \cite{barth15}. This test illustrates the current capabilities of the pipeline and the plausibility of the initial results. We also see that all of the velocity shifts are positive, which indicates that all of the profiles have been redshifted as a result of special and general relativistic effects incorporated in the line profile calculations. The observed centroid shift corresponds to the combined effect of the gravitational and transverse redshift at a radius of $\sim$ 5000 $r_g$ in the disk, which is near the middle of the assumed line-emitting annulus.

\begin{figure}[ht!]
\epsscale{0.75}
\includegraphics[width=0.5\textwidth]{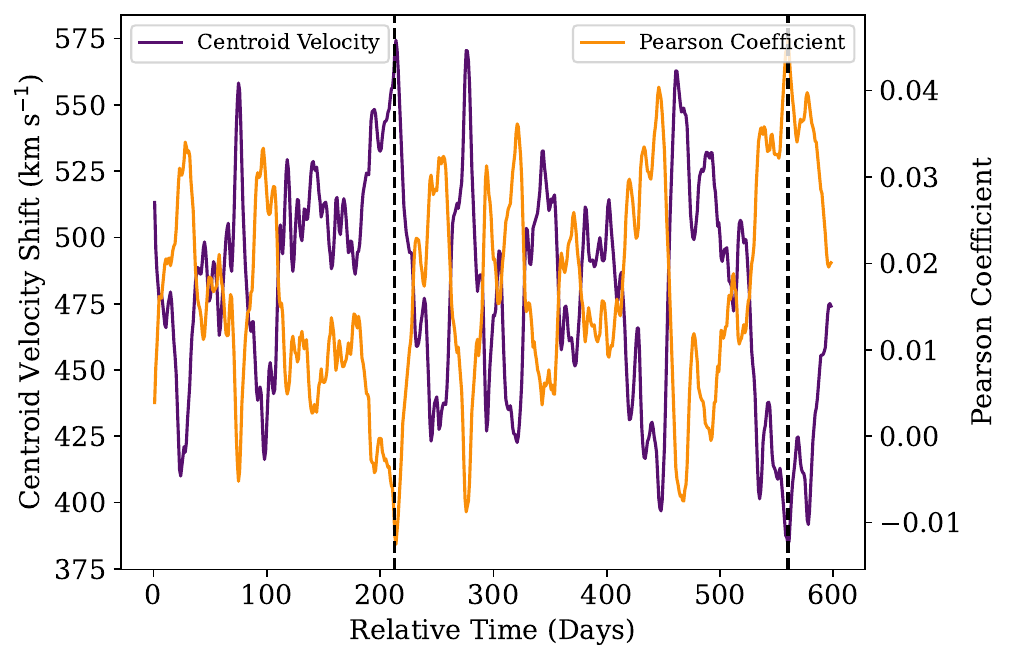}
\caption{The centroid velocity shift and Pearson skewness coefficient plotted together as a function of time, where the centroid velocity shift is in purple and the Pearson coefficient is in orange. The two vertical lines highlight points in time where example profiles were taken, as seen in Figure\ \ref{fig:cenpearprofiles}. As the inner part of the spiral arm, seen in Figure\ \ref{fig:spiralarm}, gets preferentially illuminated by reverberation, the centroid wavelength shifts to the red, as does the median wavelength due to asymmetries of the profile. Thus, there is an apparent anti-correlation between the centroid wavelength shift and the Pearson skewness coefficient. 
\label{fig:cenpear}}
\end{figure}

\begin{figure}[ht!]
\epsscale{0.75}
\includegraphics[width=0.5\textwidth]{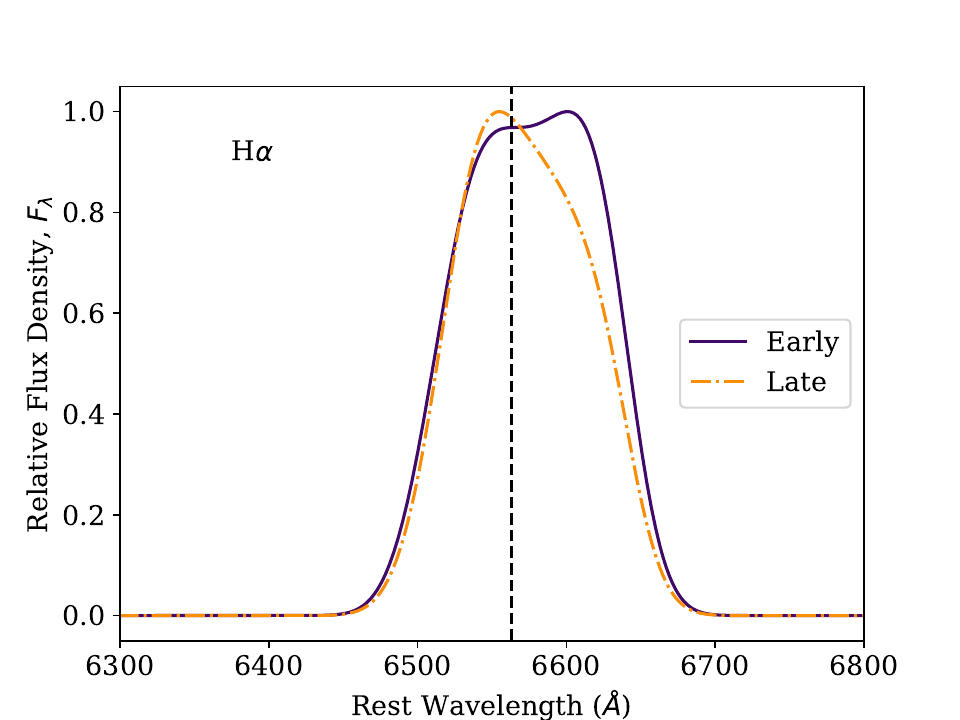}
\caption{Two example profiles taken from the test of the pipeline at times marked by the vertical dashed lines in Figure\ \ref{fig:cenpear}. Early in the time series, when the centroid wavelength is redshifted and the shape of the profile skews to the right, we see that the corresponding Pearson coefficient in Figure\ \ref{fig:cenpear} is negative due to the redshifting of the median wavelength. Later in the same time series, we see that the centroid wavelength is redshifted less and and the shape of the profile skews to the blue, and the corresponding Pearson coefficient is positive/larger due to similar smaller redshifting effects on the median wavelength of the profile. 
\label{fig:cenpearprofiles}}
\end{figure}

In the second test the disk includes a spiral arm instead of being axisymmetric and its emissivity is modified by the effects of radiative transfer. The second test utilizes the light curve presented in Figure\ \ref{fig:synlight} in black and the disk parameters from the captions of Figures\ \ref{fig:powerlaw}, \ref{fig:spiralarm}, \ref{fig:radtransfer}, with the resulting time series presented in Figure\ \ref{fig:pipelinetest}. We see a similar correlation in all three parameters that seem to be a result of the driving input light curve with a small time delay. We see that the centroid velocity shifts, presented in Figure\ \ref{fig:pipelinetest}b, are positive, which indicates an overall redshifting of the profiles. For this test, the redshifting is predominantly a result of the spiral arm and its location on the disk, which causes asymmetries in the line profiles that shift the centroid wavelength over time. Furthermore, there are similar correlated variations of the centroid velocity shifts to the continuum light curve, presented in Figure\ \ref{fig:pipelinetest}a. We see some of the effects of the non-axisymmetric features in the BLR when we examine the time series of the centroid velocity shift and Pearson skewness coefficient, which are presented in Figure\ \ref{fig:pipelinetest}b and d. Inspection of Figures\ \ref{fig:spiralarm} and \ref{fig:timedelay} shows that light echoes preferentially illuminate different portions of the spiral arm at different times, which results in evolving asymmetries of the line profiles and changes of the centroid wavelengths. This preferential illumination increases the overall skewness of the profile and produces the effect illustrated in Figure\ \ref{fig:cenpear}. By plotting profiles from the times indicated on Figure\ \ref{fig:cenpear} with the dashed vertical lines, we see the resulting asymmetries in the line profiles and how they shift with time, as illustrated in Figure\ \ref{fig:cenpearprofiles}.  

%We begin to see some of the effects of fluctuating illumination and non-axisymmetric features in the BLR. The emissivity law we used during much of the testing, $\epsilon \sim r^{-3}$, creates the correlation between velocity dispersion and the intensity of the continuum. This correlation results in moving the intensity-weighted region of the disk producing the broad line (i.e., `the center of light'). pulls the intensity-weighted region closer in, where it feels more gravitational effects and therefore more width increase towards the longer wavelength wing, which in turn reddens the centroid wavelength. This movement of the centroid wavelength to longer wavelengths increases the skewness of the emission line profile, and can be closely seen in Figure\ \ref{fig:cenpear}. The centroid shift therefore largely reflects changes in the skewness of the emission line profiles.

It is also noteworthy that the two particular configurations used in the tests lead to anti-breathing, in that a stronger illuminating continuum produces broader line profiles. This effect is seen in the apparent correlation of the velocity dispersion in Figure\ \ref{fig:pipelinetest_sym}c and Figure\ \ref{fig:pipelinetest}c to their respective light curves. 

\subsection{Summary and Future Work} \label{subsec:future}
In this paper, we have presented a pipeline that can simulate reverberation from a disk-like BLR. We use physical models and analytic prescriptions to specify features of the BLR, such as the geometry, kinematics, and emissivity. We incorporate special and general relativistic effects in the calculation of the line profiles. We focus specifically on the disk-like geometry for the BLR, but this pipeline can be adapted to use other geometries and velocity fields as well. We produce synthetic broad emission line profiles and measure their intrinsic shape parameters to see the changes they exhibit over the light-crossing time of the BLR, which may help further inform us on the BLR kinematics and geometry. Our pipeline does not require significant computational power or computational time when running one realization of the pipeline, and averages approximately less than a minute for one realization, including the saving of intermediary steps. This speed can help identify narrow parameter spaces that can be further investigated using other sophisticated forward-modeling codes introduced in Section \ref{sec:intro}.   

In order to increase the speed of this pipeline, work is currently underway to parallelize some of the algorithms. Further work will explore the full parameter space of this model in order to investigate conditions that can cause the breathing or anti-breathing behavior, as well as the velocity jitter. Our forthcoming parameter space exploration will be guided by the range in disk parameters found by previous studies (see Section \ref{subsec:emissivity}). As mentioned in Section \ref{subsec:discussion}, we will also incorporate a non-linear responsivity of the disk. Future applications will use long duration synthetic light curves that can span the dynamical time of the BLR. Using other empirical light curves, such as those from the Sloan Digital Sky Survey (SDSS) V Reverberation Mapping Campaign, can provide direct comparisons between the behaviors of our synthetic broad emission line profiles and those that are observed, while also allowing exploration of the model's parameter space. With light curves expected from future telescopes and surveys such as the Long Synoptic Survey Telescope \citep{ivezic19}, we can utilize light curves with baselines as small as 200 days up to months to interpret observed line profiles. In a forthcoming paper, we will present the results of a parameter space exploration, as well as the consequences of different combinations of parameters for velocity jitter and breathing.

\clearpage

%\section{Manuscript styles} \label{sec:style}

%The default style in \aastex\ v7 is a tight single column style, e.g. 10
%point font, single spaced.  The single column style is very useful for
%articles with wide equations. It is also the easiest style to work with
%since figures and tables, see Section \ref{sec:floats}, will span the
%entire page, reducing the need for address float sizing.

%To invoke a two column style similar to what is produced in
%the published PDF copy use: \\

%\noindent {\tt\string\documentclass[twocolumn]\{aastex7\}}. \\

%% Please use the acknowledgment and contribution environments. This will 
%% be anonomyized when the "anonymous" style option is used. 
\begin{acknowledgments}
We thank the anonymous referee for the thoughtful and constructive comments that helped us improve the paper. M.O., M.E., and N.N.M. acknowledge support from the National Science Foundation through grant AST-2205720 (WoU-MMA). M.O. and N.N.M. acknowledge support from the Pennsylvania State Science Achievement Graduate Fellowship Program. J.C.R. and C.M.D. acknowledge support from the National Science Foundation (NSF) under grant AST-2205719 and the the National Aeronautics and Space Administration (NASA) LISA Preparatory Science Program under award 80NSSC22K0748. We thank Kadri Nizam and Laura Duffy for thoughtful and technical discussions regarding the algorithmic processes. We also thank Erika Sottocorno for helpful insights, discussions, and collaborations.
\end{acknowledgments}

%% For this sample we use BibTeX plus aasjournalv7.bst to generate the

\bibliography{sample7}{}
\bibliographystyle{aasjournalv7}



\end{document}